\documentclass[reprint,aps,pra,twocolumn,showpacs]{revtex4-2}

\usepackage{graphicx}
\usepackage{dcolumn}
\usepackage{bm}
\usepackage{hyperref}
\usepackage[mathlines]{lineno}

\hypersetup{colorlinks=true, citecolor=blue, urlcolor=blue, linkcolor=blue}
\usepackage{slashed}  

\usepackage{amsmath}
\usepackage{amssymb} 
\usepackage{natbib}
\usepackage{mathrsfs}
\usepackage{flushend}
\makeatletter

\newcommand{\Rmnum}[1]{\expandafter\@slowromancap\romannumeral #1@}
\makeatother
\allowdisplaybreaks[3]

\begin{document}
\preprint{APS/123-QED}

\title{Generalized Optics-Free Cross-Correlation Ghost Imaging via Holographic Projection with Grayscale and Binary Amplitude-only Computer-Generated Holograms}



\author{Yuhan Guo}
\author{Xiangyu Yin}
\author{Chunguang Meng}
\author{Liming Li}
\email{liliming@sdut.edu.cn}
\author{Huiqiang Liu}
\email{liuhq@sdut.edu.cn}

\affiliation{School of Physics and Optoelectronic Engineering, Shandong University of Technology, Zibo 255049, China\\}


\date{\today}

\begin{abstract}
In certain applications or wavelength regimes, essential optical components for imaging systems are either unavailable or challenging to fabricate. To address this, we propose an optics-free classical ghost imaging (GI) scheme utilizing visible light. By employing grayscale and 0–1 binary amplitude-only computer-generated holograms (CGHs)—generated via a modified Gerchberg–Saxton algorithm combined with Otsu’s thresholding method—we achieve accurate replication of light intensity distributions with central symmetry in the holographic projection plane. Experimentally, we first optimized system parameters by analyzing the point spread function (PSF) and subsequently demonstrated cross-correlation GI through the precise replication of dynamic speckle patterns. Furthermore, by incorporating sparse target patterns, we significantly enhanced the imaging quality. Given the high-speed modulation capabilities of digital micromirror devices (DMDs) for 0–1 binary amplitude-only CGHs, the proposed scheme represents a significant advancement toward practical implementation, particularly in the X-ray regime where conventional optics are difficult to employ.
\end{abstract}
\maketitle

\section{Introduction}
In 1948, physicist Dennis Gabor first proposed the concept of optical holography~\cite{gabor1948new}, a contribution for which he was awarded the Nobel Prize in Physics in 1971. Subsequent research has significantly advanced this promising field~\cite{1965Spatial,tonomura1987applications,kreis1997methods,arkani2001holography}. Further progress was achieved after R. W. Gerchberg proposed the Fourier iterative method, commonly known as the Gerchberg–Saxton (GS) algorithm~\cite{Gerchberg1972APA}. It is well established that the GS algorithm has been a key factor accelerating both theoretical and experimental developments in computational holography. The GS algorithm and related improved methods~\cite{J1982Phase,guo2015iterative} provide essential theoretical frameworks for the rapid and accurate generation of computer-generated holograms (CGHs).

Holographic displays can be realized using various experimental setups, such as liquid-crystal spatial light modulators (SLMs)~\cite{zhou2021progress}, digital micromirror devices (DMDs)~\cite{hoyos2024non,zea2018optimized}, and metamaterials~\cite{2016Nonlinear,2020Binary}. Holographic-projection-based ghost imaging (GI) urgently requires high-speed modulation devices, particularly DMDs~\cite{Li2025Integration}. However, the DMD is limited to 0–1 binary amplitude-only modulation due to the tilt of pixel mirrors, which possess only dual states. By employing Otsu’s algorithm~\cite{OtsuAuto1975AThreshold}, grayscale CGHs can be efficiently and conveniently binarized~\cite{OtsuAuto1975AThreshold,2016Binary}. Thus, a holographic-projection-based GI~\cite{Li2025Integration} scheme utilizing grayscale amplitude-only CGHs can be realized using high-frame-rate DMDs following CGH binarization.

As a powerful tool for information transmission and data storage, optical imaging has evolved from conventional lens-based imaging into advanced computational imaging suited for complex scenarios, such as non-line-of-sight imaging~\cite{Katz2014Non}, imaging through scattering media~\cite{2012Non}, and nanoscale super-resolution~\cite{S1994Breaking}. As a branch of computational imaging, ghost imaging (GI) with classical light sources has been investigated for over two decades~\cite{2004Quantum,gatti2004correlated,2005Two,2005Correlated,2004Incoherent,2016Computational}. Its operational spectrum has expanded from visible light~\cite{shapiro2008computational,2009Two} to ultraviolet~\cite{2024Computational,Jin:21}, infrared~\cite{wu2024mid,2025Mid,2023A}, and even X-ray regimes~\cite{zhao2025parallel,Daniele2016Experimental,2016Fourier,2023Fourier}. However, conventional optical components may become impractical or ineffective in certain environments or specific wavelength ranges; examples include glass lenses, DMDs, and photon detectors. Consequently, GI schemes that operate without such optical elements are of significant value in frontier research~\cite{2017X,2017Table,2009Lensless,Zheng:24}.

In two recent preprints~\cite{Li2025Active,Li2025Integration}, we demonstrated that the peak value of the Hanbury Brown–Twiss effect (also known as the bunching effect) can be actively controlled using phase-only CGHs. Our results indicate that holographic projection patterns exhibit a super-bunching effect when the target pattern is either the intensity-squared thermal speckle patterns or artificially designed sparse matrix target patterns~\cite{Li2025Active}. Moreover, we showed that both types of projection patterns can enhance the visibility of ghost imaging (GI)~\cite{Li2025Integration}. Notably, when the target pattern is defined as a uniform matrix of ones, the resulting holographic projection exhibits intensity statistics identical to those of a thermal speckle pattern~\cite{Li2025Active}.

In this work, we propose an improved Gerchberg–Saxton (GS) algorithm to generate grayscale amplitude-only computer-generated holograms (CGHs), which are subsequently binarized into 0–1 binary amplitude-only patterns. To the best of our knowledge, the accurate replication of light intensity using amplitude-only CGHs has not been reported previously. Notably, such accurately replicated patterns can substantially simplify the experimental setup of ghost imaging (GI). For spectral bands—such as X-rays—where conventional optics are unavailable or impractical, fabricating 0–1 binary amplitude-only masks provides a viable route to achieve accurate light intensity replication on the target projection plane. In our proof-of-principle experiment, the accurate replication of the light intensity distribution of the reconstructed pattern is verified. Subsequently, these patterns are adopted in a GI scheme with an optics-free setup. Furthermore, the image quality of the GI scheme is enhanced by employing artificially designed sparse matrix target patterns.

\section{Theoretical model} 
The Gerchberg–Saxton (GS) algorithm and its variants represent a class of classical iterative algorithms widely applied in CGH and phase retrieval. To generate a CGH, the core concept involves alternately applying constraints between the spatial domain, ${Ae^{i{\varphi _{a}}}}$ (diffraction plane), and the frequency domain, $B{e^{i{\varphi _b}}}$ (detection plane), progressively converging toward the target phase $\varphi _{a}$. Specifically, in the diffraction plane, the amplitude $A$ is set to unity, representing a uniform amplitude at each iteration. In the detection plane, $B$ represents the target amplitude distribution serving as the constraint, while the phase $\varphi _b$ is trivial. Thus, the traditional lens-free amplitude-only CGH can be expressed as:
\begin{equation}\label{C1001}
	\begin{split}
		AO_{\mathrm{Trad}} &= \Re\left\{ e^{i\varphi_{a}} e^{-ik\vec{r}^{2}/(2f)} \right\} \\
		&=\frac{1}{2}e^{i\varphi_{a}} e^{-ik\vec{r}^{2}/(2f)} + \frac{1}{2}e^{-i\varphi_{a}} e^{ik\vec{r}^{2}/(2f)},
	\end{split}	
\end{equation}
where $\varphi _{a}$ is the locally optimal phase solution obtained via the GS algorithm, $\Re\{\cdots\}$ denotes the real-part extraction operation, $k=2\pi/\lambda$ is the wavenumber with light wavelength 
$\lambda$, $\vec{r}$ represents the two-dimensional coordinates, and $f$ is the propagation distance between the CGH plane and the detection plane. Note that the holographic reconstruction exhibits a conjugate intensity pattern originating from the Fourier transforms of the terms $\frac{1}{2}e^{i\varphi_{a}}$ and $\frac{1}{2}e^{-i{\varphi _{a}}+2ik\vec{r}^2/(2f)}$. However, the intensity of the Fourier transform of $\frac{1}{2}e^{-i{\varphi _{a}}+2ik\vec{r}^2/(2f)}$ is a unfocused noise.

However, if a perfect conjugate projection can be generated by the GS algorithm, it could be applied to  single-arm GI~\cite{Li2025Integration,LiPRA2019Transverse}. Notably, this goal can be achieved by an improved GS method utilizing amplitude-only CGHs, as follows:
\begin{equation}\label{C1002}
	\begin{split}
		AO_{\text{Imp}} &=\Re\{ \Re\{e^{i{\varphi _{a}}}\} e^{-ik\vec{r}^2/(2f)} \}\\
		&=\frac{1}{4}e^{i{\varphi _{a}}} e^{-ik\vec{r}^2/(2f)}+\frac{1}{4}e^{-i{\varphi _{a}}} e^{-ik\vec{r}^2/(2f)}\\
		&+\mathcal{N},\\
	\end{split}	
\end{equation}
where
\begin{equation}\label{C1003}
	\begin{split}
	\mathcal{N}=&\frac{1}{4}e^{i{\varphi _{a}}+2ik\vec{r}^2/(2f)} e^{-ik\vec{r}^2/(2f)}\\
	+&\frac{1}{4}e^{-i{\varphi _{a}}+2ik\vec{r}^2/(2f)} e^{-ik\vec{r}^2/(2f)}.
	\end{split}	
\end{equation}
Significantly, the first two terms in Eq.~\ref{C1002} yield an accurate replication of light intensity in a centrally symmetric manner. Although the noise term $\mathcal{N}$ persists at this stage, its adverse effects on holographic reconstruction can be avoid as adopted, as demonstrated in the subsequent experimental verification..

In this work, the binarization is performed using Otsu’s algorithm, a non-iterative global thresholding approach~\cite{OtsuAuto1975AThreshold}. Using the optimal threshold value $t^{*}$ determined by Otsu’s algorithm, each element of $AO_{\text{Imp}}$ is processed pixel-wise according to the following rule:
\begin{equation}\label{C1004}
	AO_{\mathrm{BI\text{-}Imp}} = 
	\begin{cases} 
		1, & \text{if: } AO_{\mathrm{Imp}} \geq t^{*} \\
		0, & \text{if: } AO_{\mathrm{Imp}} < t^{*},
	\end{cases}
\end{equation}
where $AO_{\text{BI-Imp}}$ denotes the resulting improved binary (0–1) amplitude-only CGH.

\section{Experimental verification}
\begin{figure}[!b]
	\centering
	\includegraphics[width=0.485\textwidth]{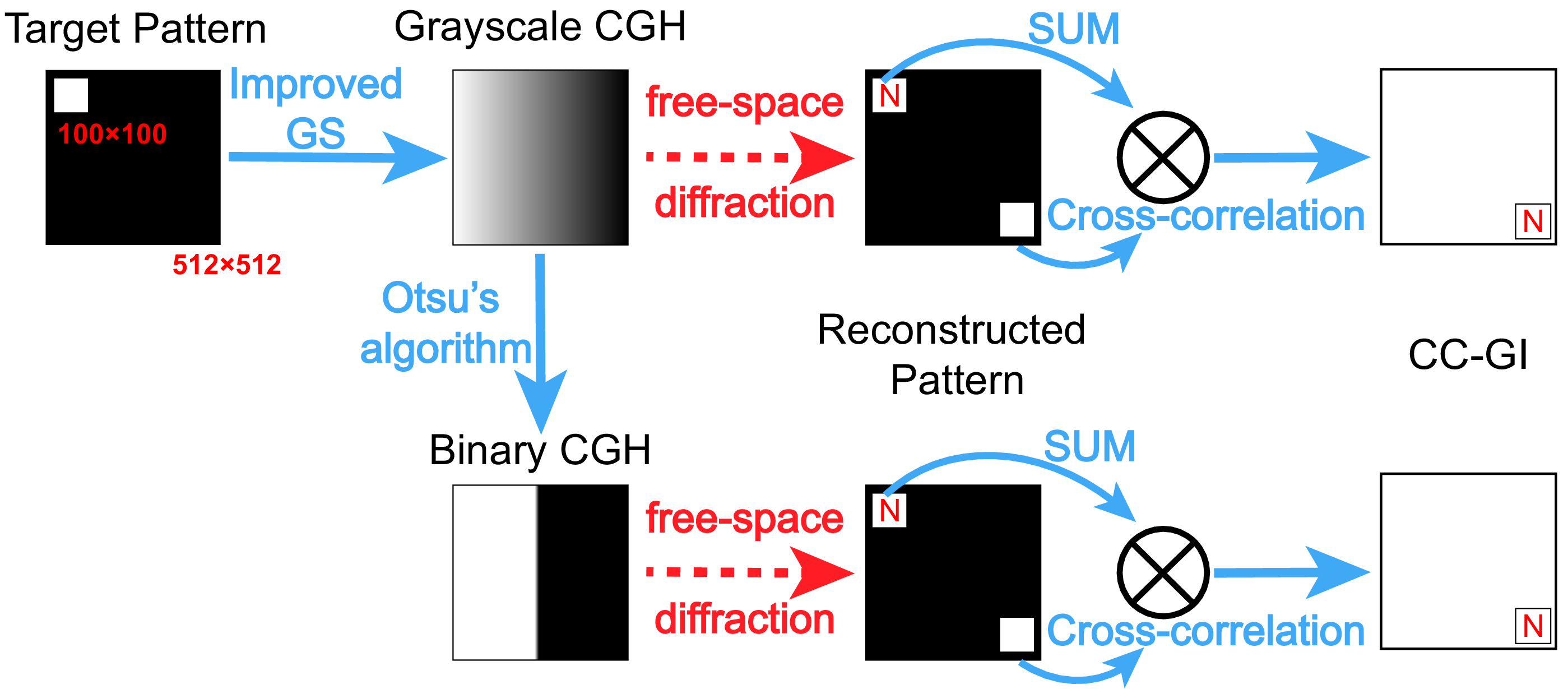}
	\caption{Data flow diagram of cross-correlation ghost imaging (CC-GI) utilizing perfect conjugate projection. Blue solid arrows represent numerical operations performed in the computer, while red dashed arrows denote free-space diffraction by CGHs in the experimental setup. GS, Gerchberg–Saxton algorithm; CGH, computer-generated hologram; SUM, light intensity summation; CC-GI, cross-correlation ghost imaging.\label{01}}
\end{figure}

Figure~\ref{01} illustrates the data flow diagram of cross-correlation ghost imaging (CC-GI) utilizing perfect conjugate projection. To mitigate the effect of the noise term $\mathcal{N}$, we constructed a target pattern with an all-zero background (512×512 pixels), embedding an all-one matrix (100×100 pixels) exclusively in the upper-left region as the effective target. Based on the improved GS algorithm proposed in our theoretical model, a grayscale CGH was generated. In the experiment, free-space diffraction of the grayscale CGH was performed using an amplitude-only SLM, resulting in a reconstructed pattern containing a perfect conjugate projection. Subsequently, the test object (the letter 'N') was placed within the effective area of the reconstructed pattern, and the transmitted light intensity was recorded. This process was repeated 10,000 times to obtain a single CC-GI result. The sum of light intensity within the effective area of the reconstructed pattern was then cross-correlated with the conjugate pattern, ultimately realizing CC-GI. Furthermore, we conducted CC-GI experiments using a binarized CGH following the exact same procedure, with the only difference being that the grayscale CGH was binarized via Otsu's algorithm.

\begin{figure}[!t]
	\centering
	\includegraphics[width=0.485\textwidth]{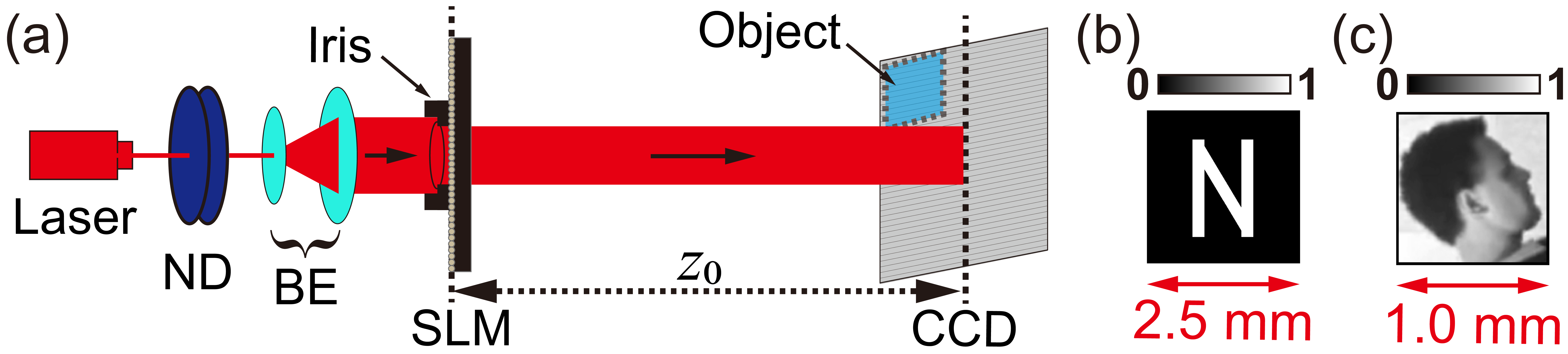}
	\caption{(a) Schematic of the experimental setup for cross-correlation ghost imaging with holographic projection. ND, neutral density filter; BE, beam expander; SLM, amplitude-only spatial light modulator; CCD, charge-coupled device camera. The test objects in (b) and (c) are binary and grayscale, respectively.\label{02}}
\end{figure}

Figure~\ref{02} illustrates the experimental setup for CC-GI with holographic projection, along with the two test objects. In Fig.~\ref{02}(a), a single-mode continuous-wave laser beam ( $\lambda$ = 632.8 nm) is attenuated, collimated, and shaped using two neutral density (ND) filters and a beam expander (BE). The modulated beam passes through an iris and an amplitude-only SLM (an element pixel size of $18\times18$ $\mu$$\text{m}^2$ and a total pixels $1024\times768$, custom-made, Shandong Normal University, China). It subsequently undergoes free-space diffraction before being captured by a charge-coupled device (CCD) camera (an element pixel size of $5.86\times5.86$ $\mu$$\text{m}^2$ and a total pixels $1920\times 1200$, MER2-231-41U3M, Daheng Optics, China). A test object, indicated as a light-blue square in Fig.~\ref{02}(a), is placed in direct contact with the CCD detector. The axial distance between the SLM and CCD planes is $ z_0 = 0.4 $ m. The beam diameter $ D $ is adjusted by the iris, which is positioned as close as possible to the SLM. Figures~\ref{02}(b) and \ref{02}(c) display a binary test object (the character 'N') and a grayscale test object (a portrait), respectively.

\subsection{Perfect conjugate projection and Point spread function}
\begin{figure}[!hbtp]
	\centering
	\includegraphics[width=0.40\textwidth]{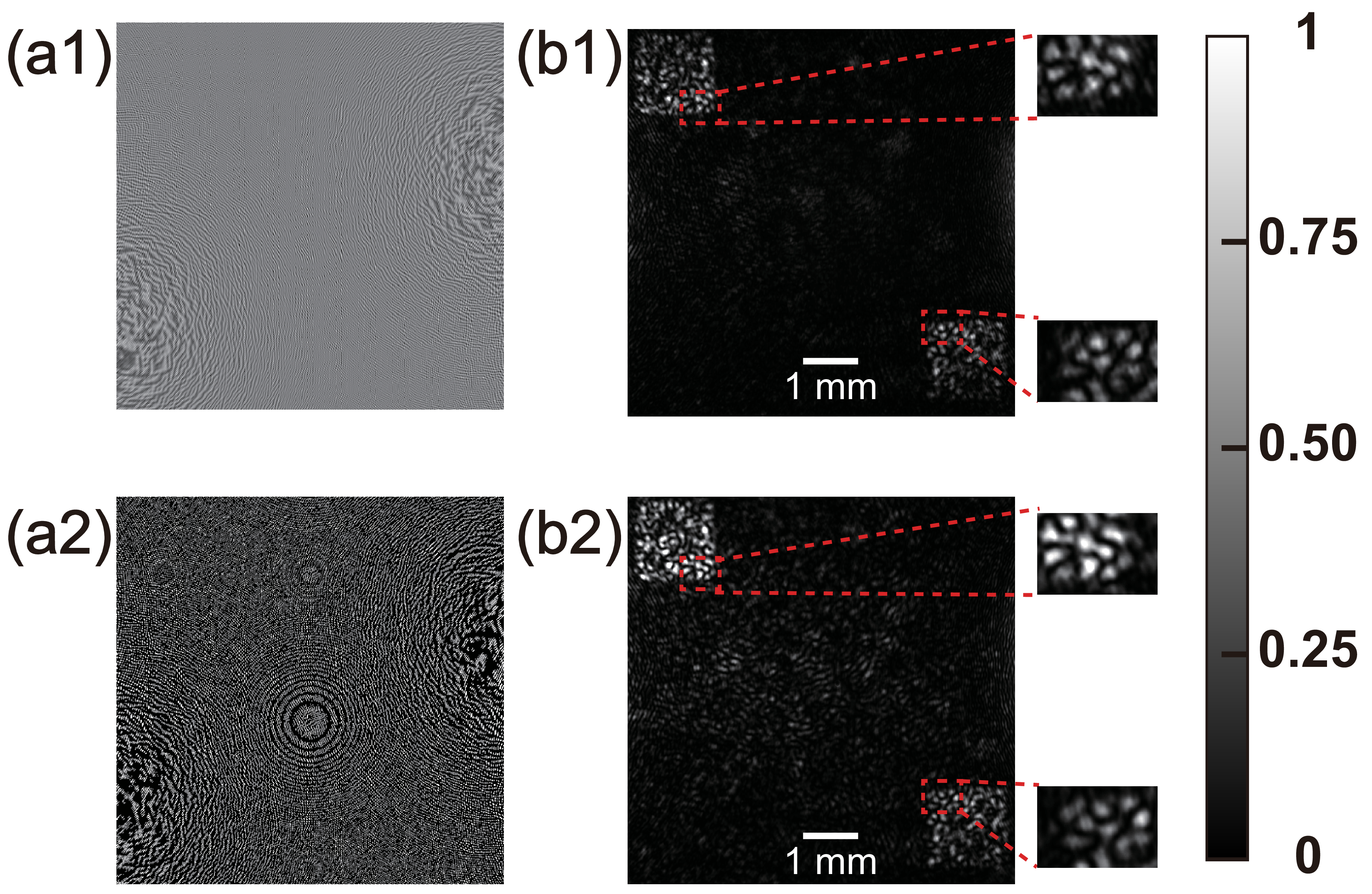}
	\caption{Key results of holographic projection using the target pattern shown in Fig.~\ref{01}. (a1) Grayscale amplitude-only CGH. (a2) Binary (0–1) CGH derived from (a1) via Otsu’s algorithm. (b1) and (b2) Reconstructed patterns corresponding to the CGHs in (a1) and (a2), respectively. The insets on the right of (b1) and (b2) demonstrate accurate replication of light intensity with central symmetry. All four main subgraphs and their insets share a common color bar.\label{03}}
\end{figure}

We removed the test object to demonstrate the perfect conjugate projection generated by the improved GS algorithm proposed in our study. Figure~\ref{03} presents the key results of the computational holography. The target pattern, illustrated in Fig.~\ref{01}, has been described in detail previously and thus is not reiterated here. Figure~\ref{03}(a1) displays the grayscale amplitude-only CGH obtained using the proposed improved GS algorithm. Using Otsu’s algorithm~\cite{OtsuAuto1975AThreshold}, the grayscale CGH was binarized to generate the 0–1 binary CGH shown in Fig.~\ref{03}(a2). Furthermore, Figs.~\ref{03}(b1) and \ref{03}(b2) depict the reconstructed patterns corresponding to the grayscale and 0–1 binary CGHs in Figs.~\ref{03}(a1) and \ref{03}(a2), respectively. As discussed in our previous work~\cite{Li2025Active}, the speckle features observed in these reconstructed patterns are a direct result of the intrinsic properties of optical interference. The zoomed insets on the right of Figs.~\ref{03}(b1) and \ref{03}(b2) demonstrate an accurate replication of light intensity with central symmetry. Although the background noise in Fig.~\ref{03}(b2) is more pronounced than in the reconstructed pattern shown in Fig.~\ref{03}(b1), the speckle patterns across all four insets exhibit remarkable similarity, despite the limited precision of the custom-made SLM.

To evaluate the application potential of the perfect conjugate projection, we systematically investigated the influence of two key parameters—the hologram diameter $D_{\text{CGH}}$ and the number of pixels $N$ along one dimension of the effective target pattern—on the point spread function (PSF) of the GI scheme. In this scheme, the PSF is equivalent to the normalized second-order correlation function, $g^{(2)}(x_1-x_2)$, of the speckle pattern. Consequently, the peak value $g^{(2)}(0)$ serves as a direct metric for assessing the GI performance of the perfect conjugate projection.

\begin{figure*}[!hbtp]
	\centering
	\includegraphics[width=0.60\textwidth]{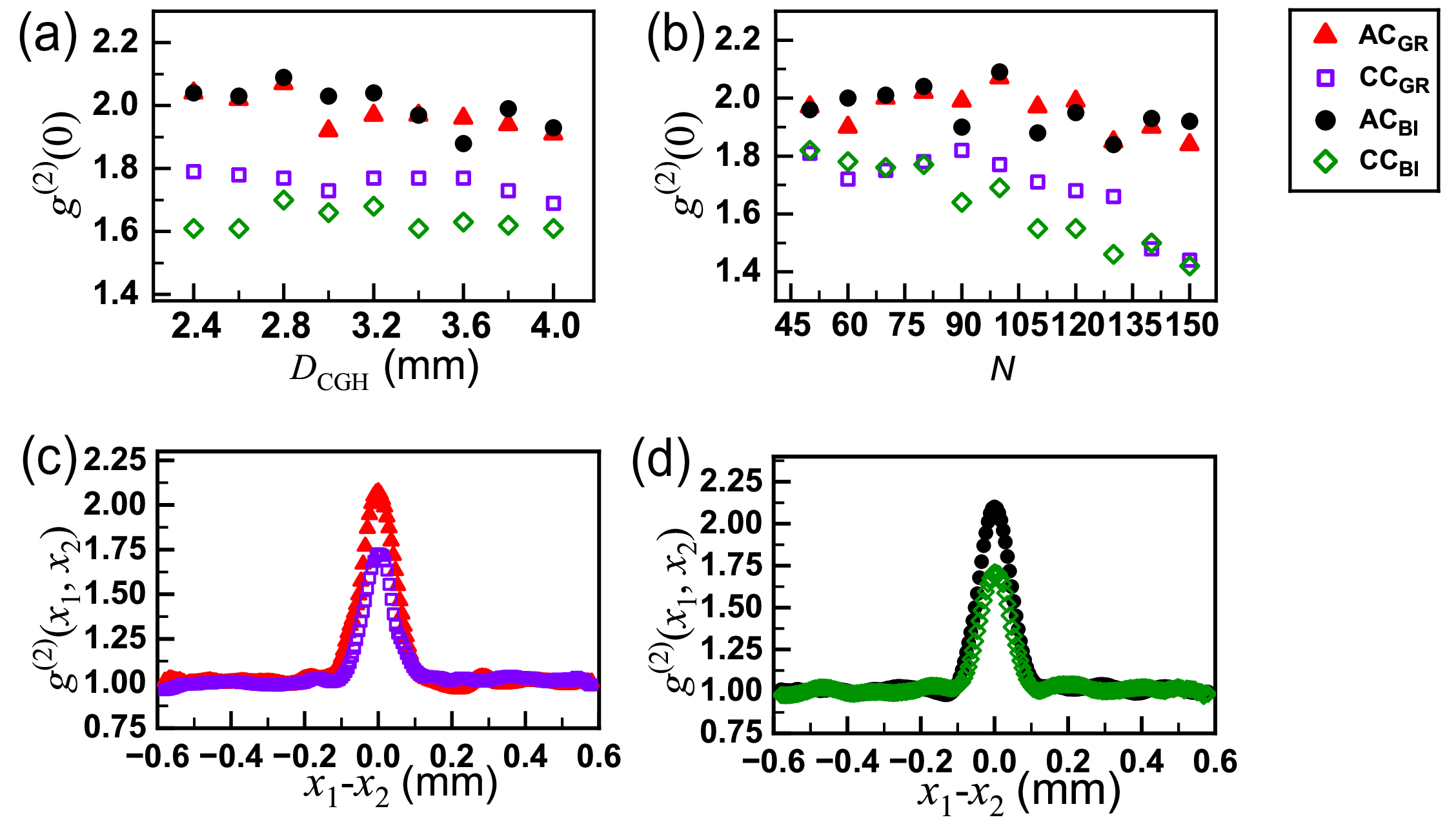}
	\caption{Experimental results of the normalized second-order correlation function for holographic reconstructed patterns using an all-ones matrix as the effective target pattern. (a) Relationship between $g^{(2)}(0)$ and the CGH diameter $D_{\text{CGH}}$. Here, the effective target pattern consists of $N$ = 100 pixels per dimension. (b) Relationship between $g^{(2)}(0)$ and the number of pixels $N$ per dimension in the effective target pattern. Here, the CGH diameter $D_{\text{CGH}}$ is 2.8 mm. (c) and (d) Second-order bunching curves for holographic reconstructed patterns generated from grayscale and binary amplitude-only CGHs, respectively, with parameters $D_{\text{CGH}}$ = 2.8 mm and $N=100$. The measured values for the auto-correlation ($\text{AC}_{\text{GR}}$, red hollow triangles) and cross-correlation ($\text{CC}_{\text{GR}}$, purple hollow squares) cases using reconstructed patterns from grayscale CGHs are indicated. Similarly, the corresponding values obtained from binary CGH reconstructions are marked by black hollow circles ($\text{AC}_{\text{BI}}$) and green hollow diamonds ($\text{CC}_{\text{BI}}$), respectively.\label{04}}
\end{figure*}

Here, we further investigate the influence of two key parameters of the holographic projection system, the CGH diameter $D_{\text{CGH}}$ and the number of pixels $N$ per dimension in the effective target pattern, on the peak value $g^{(2)}(0)$. Figure~\ref{04} presents the experimental results of the normalized second-order correlation function $g^{(2)}(x_1,x_2)$ for holographic reconstructed patterns generated using an all-ones matrix as the effective target pattern. 

\begin{figure*}[!htpb]
	\centering
	\includegraphics[width=0.75\textwidth]{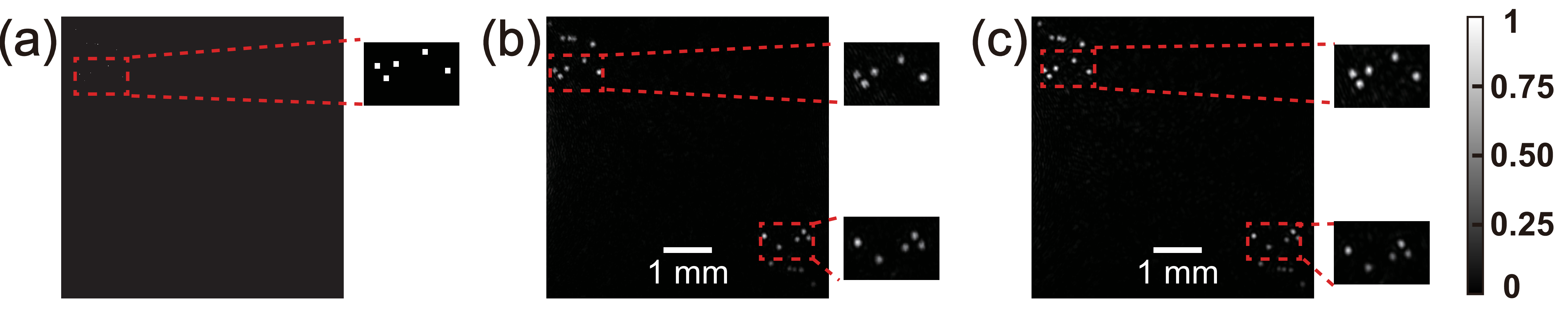}
	\caption{Sparse matrix target pattern and holographic reconstructed patterns. (a) Target pattern: an all-zeros matrix with $0.1\%$ of the pixels randomly set to 1 in the upper-left 100×100 region. (b) and (c) Holographic reconstructed patterns generated from grayscale and binary (0–1) amplitude-only CGHs, respectively. All three subplots and the five insets share a common color bar.\label{05}}
\end{figure*}

Figure~\ref{04}(a) plots the peak value $g^{(2)}(0)$ against the diameter $D_{\text{CGH}}$ with $N=100$. Considering both grayscale and binary CGH scenarios, four distinct $g^{(2)}(0)$ values are obtained for each diameter: (i) auto-correlation from grayscale CGH reconstructions ($\text{AC}_{\text{GR}}$, red hollow triangles); (ii) cross-correlation from grayscale CGH reconstructions ($\text{CC}_{\text{GR}}$, purple hollow squares); (iii) auto-correlation from binary CGH reconstructions ($\text{AC}_{\text{BI}}$, black hollow circles); and (iv) cross-correlation from binary CGH reconstructions ($\text{CC}_{\text{BI}}$, green hollow diamonds). In these measurements, the first detection point is fixed at the center of the holographic reconstructed pattern corresponding to the effective target pattern. The second point is either set to coincide with the first (yielding auto-correlation) or positioned at the conjugate location of the first point (yielding cross-correlation).

It should be emphasized that all $g^{(2)}(x_1,x_2)$ data presented in this work are derived from correlation calculations based on 10,000 intensity patterns of perfect conjugate projection. Since the GS algorithm tends to converge to local optima, we performed 10,000 independent restarts for the single effective target pattern (i.e., the all-ones matrix shown in Fig.~\ref{01}). Each restart yielded a corresponding grayscale CGH after 50 iterations, which was then binarized using Otsu's thresholding algorithm to obtain the binary CGH.

As observed, the maximum $g^{(2)}(0)$ value for the $\text{CC}_{\text{BI}}$ case is achieved at $D_{\text{CGH}} = 2.8$~mm. Furthermore, Figure~\ref{04}(b) shows the peak value $g^{(2)}(0)$ as a function of $N$ with $D_{\text{CGH}} = 2.8$~mm. Notably, as $N$ increases, the peak value remains nearly constant in the auto-correlation case but decreases significantly in the cross-correlation case. Additionally, for $N=100$ and $D_{\text{CGH}} = 2.8$~mm, figures.~\ref{04}(c) and \ref{04}(d) present the bunching effect curves obtained from grayscale and binary CGHs, respectively. All four subplots in Fig.~\ref{04} share the same symbol convention, as indicated in the upper-right inset. Overall, the auto-correlation results consistently outperform the cross-correlation results, while the binarization of CGHs has a negligible effect on the peak value $g^{(2)}(0)$. To achieve a larger field of view and superior visibility, we selected $N=100$ and $D_{\text{CGH}} = 2.8$~mm for subsequent experiments.

To improve the signal-to-noise ratio of $g^{(2)}(x_1, x_2)$ and achieve better visibility for CC-GI, we conducted further validation of perfect conjugate projection using a sparse matrix target pattern designed in our previous preprints~\cite{Li2025Active,Li2025Integration}. Figure~\ref{05} presents the key results, with all subplots and insets sharing a common color bar. Here, the all-ones matrix in the effective target pattern is replaced by a sparse matrix, as shown in Fig.~\ref{05}(a). This sparse matrix is generated from an all-zeros matrix by randomly setting $0.1\%$ of the pixels to 1. Figures~\ref{05}(b) and \ref{05}(c) show the holographic reconstructed patterns generated from grayscale and binary (0-1) amplitude-only CGHs, respectively. These amplitude-only CGHs (not presented here) were created using our proposed improved GS algorithm and Otsu's thresholding algorithm. Furthermore, the zoomed insets on the right side of the subplots demonstrate perfect conjugate projection. A comparison between the projection results in Figs.~\ref{03} and \ref{05} reveals that holographic projection with the sparse target pattern exhibits a superior object-image relationship, as previously demonstrated in our preprint~\cite{Li2025Active}.

\begin{figure}[!t]
	\centering
	\includegraphics[width=0.485\textwidth]{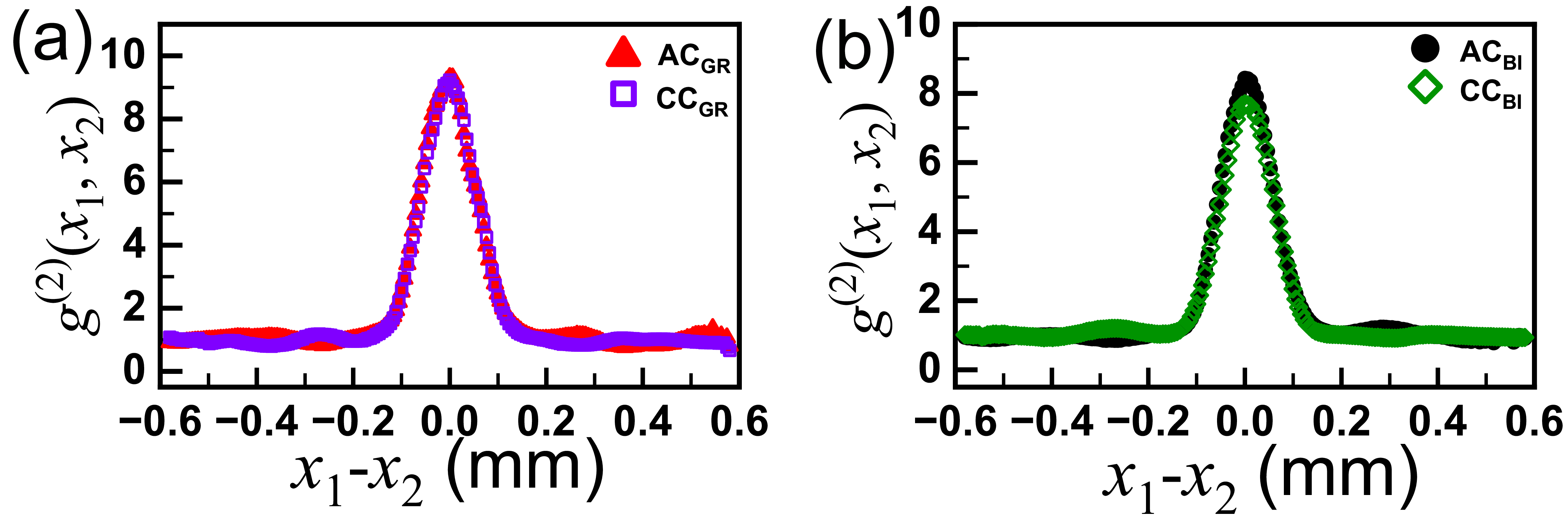}
	\caption{Bunching effect curves of holographic projection patterns generated from grayscale (a) and binary (b) amplitude-only CGHs with sparse matrix target patterns. The data visualization methods in (a) and (b) are identical to those in Figs.~\ref{04}(c) and \ref{04}(d), respectively.\label{06}}
\end{figure}

For the sparse matrix target pattern, the GS algorithm establishes a stable object–image correspondence. Consequently, we generated 10,000 random sparse matrix target patterns. Each pattern was used to compute a corresponding grayscale CGH via the improved GS algorithm. Subsequently, diffraction reconstruction experiments were performed on both the original grayscale CGHs and their binary counterparts—obtained using Otsu’s method—to acquire the corresponding reconstructed patterns for further correlation analysis. Figure~\ref{06} shows the bunching effect curves of the holographic projection patterns. Here, the positions of the two detection points used for intensity correlation in Figs.~\ref{06}(a) and \ref{06}(b), as well as the data visualization method, are identical to those in Figs.~\ref{04}(c) and \ref{04}(d), respectively; therefore, they are not described again. These results demonstrate that holographic projection with sparse matrix patterns not only significantly enhances the signal-to-noise ratio (SNR) of $g^{(2)}(x_1-x_2)$ but also ensures that the binarization of the grayscale CGH does not lead to a notable reduction in the peak value at $x_1=x_2$.

\subsection{Cross-correlation ghost imaging (CC-GI)}
\begin{figure}[!t]
	\centering
	\includegraphics[width=0.35\textwidth]{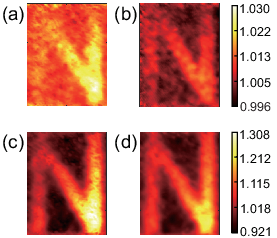}
	\caption{Cross-correlation ghost imaging results using the 0–1 binary object of the character "N" shown in Fig.~\ref{02}(b). (a, b) Imaging results for grayscale and binary amplitude-only CGHs, respectively, where the effective target pattern is an all-ones matrix. (c, d) Similarly, imaging results for grayscale and binary amplitude-only CGHs, respectively, but with 10,000 random sparse matrix target patterns. The subgraphs horizontally arranged share a common color bar.\label{07}}
\end{figure}

\begin{figure}[!b]
	\centering
	\includegraphics[width=0.35\textwidth]{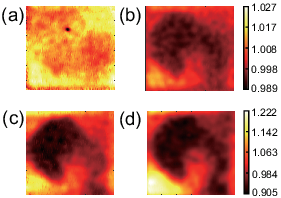}
	\caption{Cross-correlation ghost imaging results for the grayscale portrait object shown in Fig.~\ref{02}(c). The experimental parameters for subplots (a)-(d) are identical to those in the corresponding subplots of Fig.~\ref{07}. The subgraphs horizontally arranged share a common color bar.\label{08}}
\end{figure}

The perfect conjugate projection patterns described above were then applied to CC-GI. Since the CC-GI process is illustrated in Fig.~\ref{01}, it is not discussed in detail here. Figure~\ref{07} shows the CC-GI results obtained using the 0-1 binary object of the character "N" (shown in Fig.~\ref{02}(b)). The subgraphs horizontally arranged share a common color bar. Figures~\ref{07}(a) and \ref{07}(b) present the imaging results corresponding to grayscale and binary amplitude-only CGHs, respectively, where the effective target pattern is an all-ones matrix. Furthermore, Figs.~\ref{07}(c) and \ref{07}(d) display the results for grayscale and binary amplitude-only CGHs, respectively, utilizing 10,000 random sparse matrix target patterns. In the GI scheme, the visibility $V$, defined as $V=(g_{\text{max}}^{(2)}-g_{\text{min}}^{(2)})/(g_{\text{max}}^{(2)}+g_{\text{min}}^{(2)})$, depends on the maximum ($g_{\text{max}}^{(2)}$) and minimum ($g_{\text{min}}^{(2)}$) values of the second-order correlation function of the reconstructed image~\cite{Li2025Integration}. The calculated visibilities for the ghost images in Figs.~\ref{07}(a)-\ref{07}(d) are 0.0121, 0.0100, 0.1735, and 0.1404, respectively. These results demonstrate that perfect conjugate projection not only successfully enables CC-GI but also significantly enhances its visibility by leveraging sparse matrix target patterns. Additionally, although binarization has an impact on the results, it is not significant.

Furthermore, figure~\ref{08} presents the CC-GI results for the grayscale portrait object shown in Fig.~\ref{02}(c). All subplots horizontally arranged share a common color bar. Note that Fig.~\ref{08} employs the same experimental parameters as Fig.~\ref{07} for each corresponding subplot. The visibilities of the ghost images in Figs.~\ref{08}(a)-\ref{08}(d) are 0.0018, 0.0116, 0.1367, and 0.1467, respectively. These results demonstrate that employing sparse matrix target patterns enables effective CC-GI reconstruction of grayscale objects, while binary amplitude-only CGHs further enhance the ghost image quality. This enhancement is likely attributed to the superior modulation accuracy of our custom-made SLM in binary mode compared to grayscale mode.

\section{Discussion}
\subsection{Perspectives on optics-free computational imaging}
From a system-level perspective, the ND filter and BE system depicted in Fig.~\ref{02}(a) function as ancillary light source components rather than essential imaging optics. The amplitude-only SLM, operating in a 0–1 binary mode, acts as a high-density array of optical switches comprising tens of thousands of independently controllable elements. Consequently, the entire setup can be characterized as an "optics-free" imaging architecture, devoid of traditional lenses or complex optical components.

The inclusion of a BE system is necessitated by the intrinsic properties of the He–Ne laser: its output beam width is approximately one-tenth that of the CGHs, thus requiring spatial expansion to uniformly illuminate the SLM. However, this requirement is not universal. When adapting the CC-GI framework to the X-ray regime, the scenario changes significantly. The typical beam waist radius of a free-electron laser in the X-ray band inherently matches the characteristic dimensions of nanoscale CGHs, thereby obviating the need for additional beam-shaping optics~\cite{2016Fourier,pratsch2022x}.

Furthermore, the 0-1 binary amplitude modulation employed herein functions exclusively as an on/off switch, effectively circumventing the bandwidth constraints that limit conventional DMDs. More critically, by leveraging advanced nanofabrication techniques (e.g., gold- or nickel-based etching) to directly pattern 0-1 binary X-ray CGHs, the system achieves complete independence from traditional
optics~\cite{pratsch2022x,amardeep2022,william2022}. Such a minimalist configuration requires only two components: a custom 0-1 binary hologram and a photodetector tuned to the X-ray source's spectrum. In contrast to traditional pinhole imaging, our proposed CC-GI scheme constitutes a dynamic, multi-aperture, single-pixel imaging system~\cite{Katz2014Non}.

\subsection{Benefits of 0–1 binary amplitude-only modulation}
0–1 binary amplitude-only modulation exhibits distinct advantages in high-speed operation: modulators based on DMDs and FPGA-driven programmable source (e.g., LED arrays) both achieve modulation frame rates ranging from tens of kilohertz (kHz) to tens of megahertz (MHz)~\cite{Wang2017High,Peter2006High,Akbulut11}. However, DMDs are inherently constrained by the mechanical flipping mechanism of micromirrors and their micrometer-scale feature sizes, which not only limit further enhancement of modulation speed but also render them incompatible with short-wavelength regimes such as extreme ultraviolet (EUV) and X-ray regime, no commercially available DMD devices currently exist for the X-ray band~\cite{Shi2023Progress}. Notably, to address the pressing demand for high-speed, high-precision modulation in X-ray regime, researchers have successfully developed dedicated 0–1 binary amplitude-only modulation devices (e.g., nanofabricated binary masks or CGHs)~\cite{2023Fourier,DINH2019Maskless}. This advancement provides critical technical support for transcending the constraints of traditional optics and enabling CC-GI with X-ray source.

Owing to their distinct advantages of short wavelength, strong penetration capability, and high measurement precision, X-rays have been widely adopted in fields such as medical imaging~\cite{2017Table,Zhang2022Efficient}. However, the potential risk of radiation damage remains a critical bottleneck constraining their clinical and research applications. In response, low-dose X-ray ghost imaging has emerged~\cite{Zhang2022Megapixel}; recent laboratory studies have demonstrated that this technique can reduce the radiation dose required for a single image to one-thousandth of that needed by conventional X-ray imaging methods. Nevertheless, existing X-ray structured light modulation techniques—such as those employing artificial scatterers like crystal powders or porous gold films—typically generate only random light fields resembling thermal speckle. The inherent lack of controllability and coherence in these fields severely limits further improvements in imaging resolution and SNR~\cite{Zhang2022Efficient}.

To address this bottleneck, this work proposes a novel modulation strategy based on 0–1 binary amplitude-only CGHs. In contrast to traditional random scattering mechanisms, this approach fully leverages the coherence properties of source. By employing precisely engineered 0-1 binary amplitude-only structures to govern the optical diffraction process, it constructs structured fields featuring the sub-Rayleigh speckle~\cite{Yaron2014Generating}. Dominated by coherent diffraction, this light field distribution  exhibits high programmability and repeatability. Consequently, it provides a high-quality illumination paradigm for X-ray GI, thereby significantly enhancing the visibility and spatial resolution of reconstructed ghost images.

\subsection{Benefits of ghost imaging with holographic projection}
Optical projection based on CGHs represents a pivotal technology, offering a lensless solution for precisely projecting artificially designed target patten onto specified planes. The thermal light bunching effect, which serves as the physical foundation of GI, can be actively modulated via CGH: this approach enables not only the adjustment of the bunching effect's peak intensity but also the control of its peak width, thereby allowing for the flexible optimization of GI characteristics~\cite{Li2025Integration}. Furthermore, within this holographic projection framework, five distinct GI modes can be simultaneously realized~\cite{Li2025Active}, effectively integrating the advantages of two typical single-pixel imaging methodologies (SLM-SPI and DMD-SPI).

Although the generation of CGHs is computationally intensive, all calculations are performed offline prior to projection, thereby ensuring no disruption to the real-time GI measurement process. Building on this foundation, the strategy proposed in this work not only successfully generates perfect conjugate projection patterns and significantly simplifies the GI setup, but also establishes a viable and practical technical pathway for implementing GI in the X-ray regime.

\section{Conclusion} 
This paper proposes a generalized optics-free GI scheme based on a modified GS algorithm, which has been rigorously derived theoretically and validated experimentally. Under free-space diffraction conditions, an amplitude-only grayscale CGH generates a projection pattern that exhibits perfect intensity conjugation. Furthermore, results confirm that this conjugate intensity property is preserved in binary CGHs following binarization via Otsu’s algorithm. By leveraging holographic projection, the proposed approach successfully realizes an optics-free cross-correlation bunching effect and enables CC-GI using 0–1 binary amplitude-only CGHs. Moreover, the image visibility in CC-GI is significantly enhanced when a series of sparse matrices are employed as effective target patterns. Finally, this study systematically discusses the feasibility and application potential of extending holographic projection-based GI to the X-ray spectral regime.

\begin{acknowledgments}
This work was supported by National Natural Science Foundation of China (Grants No. 62105188, 12175127), Natural Science Foundation of Shandong Province, China (Grant No. ZR2025MS38) and the Scientific Innovation Project for Young Scientists in Shandong Provincial Universities (2024KJG011).
\end{acknowledgments}

\bibliography{References227llm.bib}
\end{document}